\documentclass{svproc}

\usepackage[numbers]{natbib}
\usepackage{url}

\usepackage{wrapfig}
\usepackage{hyperref}
\usepackage[version=3]{mhchem}
\usepackage{xcolor}
\usepackage{graphicx}
\usepackage{tabularx}
\newcolumntype{M}[1]{>{\centering\arraybackslash}m{#1}}

\usepackage[normalem]{ulem}

\begin{document}
\mainmatter              

\title{Studying the interaction between VUV photons and PAHs in relevant astrophysical conditions}

\titlerunning{PIRENEA and PIRENEA 2 : simulating the photochemistry of astro-PAHs}  

\author{A. Marciniak\inst{1,2,}\thanks{\email{marciniak@irsamc.ups-tlse.fr}} 
\and  A. Bonnamy\inst{1} \and  S. Zamith\inst{2} \and  L. Nogu\`es\inst{1} \and  O. Coeur-Joly\inst{1} \and  D. Murat\inst{1} \and  P. Moretto-Capelle\inst{2} \and  J.-M. L’Hermite\inst{2} \and  G. Mulas\inst{1,3} \and  C. Joblin\inst{1}}

\authorrunning{A. Marciniak et al.} 

\tocauthor{A. Marciniak, A. Bonnamy, S. Zamith,  L. Nogues,  O. Coeur-Joly, D. Murat, P. Moretto-Capelle,  J.-M. L’Hermite, G. Mulas,  C. Joblin}
\institute{Institut de Recherche en Astrophysique et Planétologie, Université Paul Sabatier - Toulouse III, CNRS, CNES, 9 Av. du Colonel Roche, F-31028 Toulouse Cdx 4, France\\
\and
Laboratoire Collisions Agrégats Réactivité (LCAR/FERMI), Université Paul Sabatier - Toulouse III, CNRS, 118 Route de Narbonne, F-31062 Toulouse, France
\and
Istituto Nazionale di Astrofisica – Osservatorio Astronomico di Cagliari, Via della Scienza 5, I-09047 Selargius (CA), Italy
}
\maketitle              

\begin{abstract}
PIRENEA and PIRENEA 2 are experimental setups dedicated to the study of fundamental molecular processes involving species of astrochemical interest. The coupling of a VUV source to PIRENEA has allowed us to study the fragmentation pathways and stability of polycyclic aromatic hydrocarbons (PAHs) containing aliphatic bonds under conditions relevant for astrophysical  photodissociation regions. PIRENEA 2 will open the possibility to extend these studies to larger systems such as PAH clusters, and more generally to study gas-nanograin-photon interactions.


\end{abstract}

\section{Introduction}
Studying the interaction of vacuum ultraviolet (VUV) photons with polycyclic aromatic hydrocarbon (PAH) molecules is of high interest to understand the physics and chemistry of photodissociation regions (PDRs) associated with star forming regions (see A.G.G.M. Tielens in this volume). This interaction is responsible for the well-known aromatic infrared band (AIB) emission but it can also induce other mechanisms such as fragmentation (see e.g., Ref~\cite{montillaud2013}). 

\begin{figure*}[!ht] 
    \centering
    \includegraphics[width=0.9\textwidth]{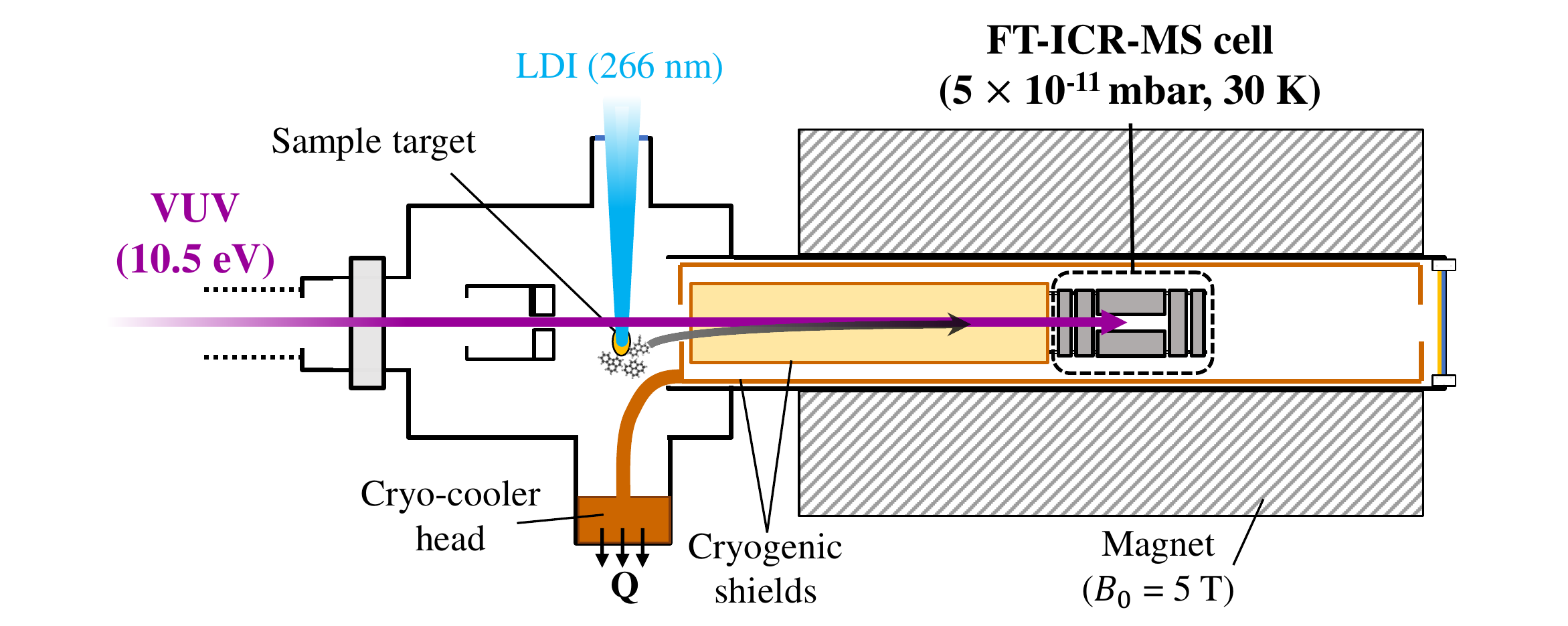}
    \caption{Scheme of PIRENEA. PAH$^+$ species are produced from a sample target by laser desorption and ionization at 266\,nm. They are stored in the cryogenic ion cyclotron resonance (ICR) cell where they are irradiated by the VUV laser. The photoproducts are then measured by non-destructive Fourier transform mass spectrometry.}
    \label{fig:PIRENEA}
\end{figure*}

PIRENEA, from the acronym of "Pi\`ege \`a Ions pour la Recherche et l'Etude de Nouvelles Esp\`eces Astrochimiques"\cite{joblin2002}, is a dedicated laboratory astrophysics setup which was developed to meet the following functionalities:
(i) isolation of molecular species of astrophysical interest for minutes in a cold ($\leq 30$\,K) and collision-free ($\leq 10^{-11}$\,mbar) environment, (ii) photoprocessing with photons in the visible to VUV range, and (iii) ultra high resolution non-destructive Fourier transform mass spectrometry (FT-MS) to record fragmentation products. The functioning principle of PIRENEA is shown and described in Figure~\ref{fig:PIRENEA}.
Here, we illustrate the ability of PIRENEA to study the VUV photoprocessing of aliphatic PAHs \cite{marciniak2021}. We put these results in perspective of future studies with the second generation setup, PIRENEA~2.

\section{VUV photo-stability of aliphatic PAHs}
\begin{wrapfigure}{r}{0.5\textwidth} 
  \centering
  \vspace{-0.4cm}
  \includegraphics[width=0.5\textwidth]{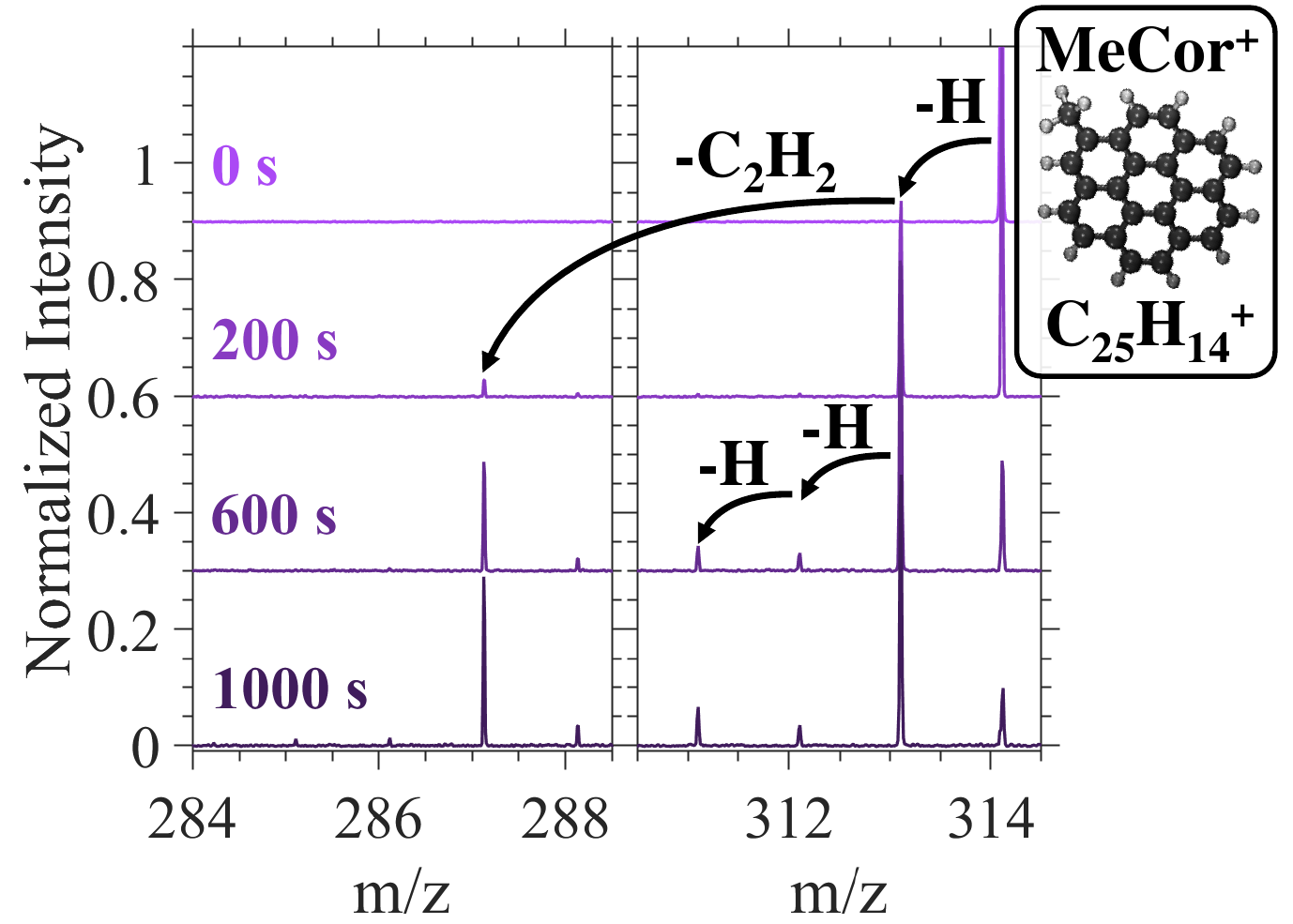}
    \caption{Mass spectra at four different VUV irradiation times for MeCor$^+$. This illustrates the fragmentation cascade induced by the VUV photoprocessing. Adapted from Ref.\cite{marciniak2021}.}    \label{fig:TOF_vs_tirr} 
\end{wrapfigure}

PAHs containing aliphatic C-H bonds are considered as good candidates for the 3.4\,$\mathrm{\mu}$m AIB observed in PDRs \cite{joblin1996}. We have recently studied the chemical evolution of aliphatic PAHs under VUV irradiation \cite{marciniak2021} by coupling a 10.5\,eV photon source to PIRENEA (Figure~\ref{fig:PIRENEA}).
This source has a low photon flux ($\sim$\,$10^{13}$\,photon~s$^{-1}$~cm$^{-2}$), which means that only one-photon processes are considered in this experiment. Besides, ions or their fragments have a sufficient time ($\sim$\,10$^2$\,s) between two photon absorption to relax most of their internal energy by radiative cooling \cite{joblin2020}. 
Thus, this setup allows the study of fragmentation cascades (see Figure~\ref{fig:TOF_vs_tirr}) under conditions that are relevant for PDRs \cite{joblin2020}.

\begin{figure*}[!ht] 
    \centering
    \includegraphics[width=1\textwidth]{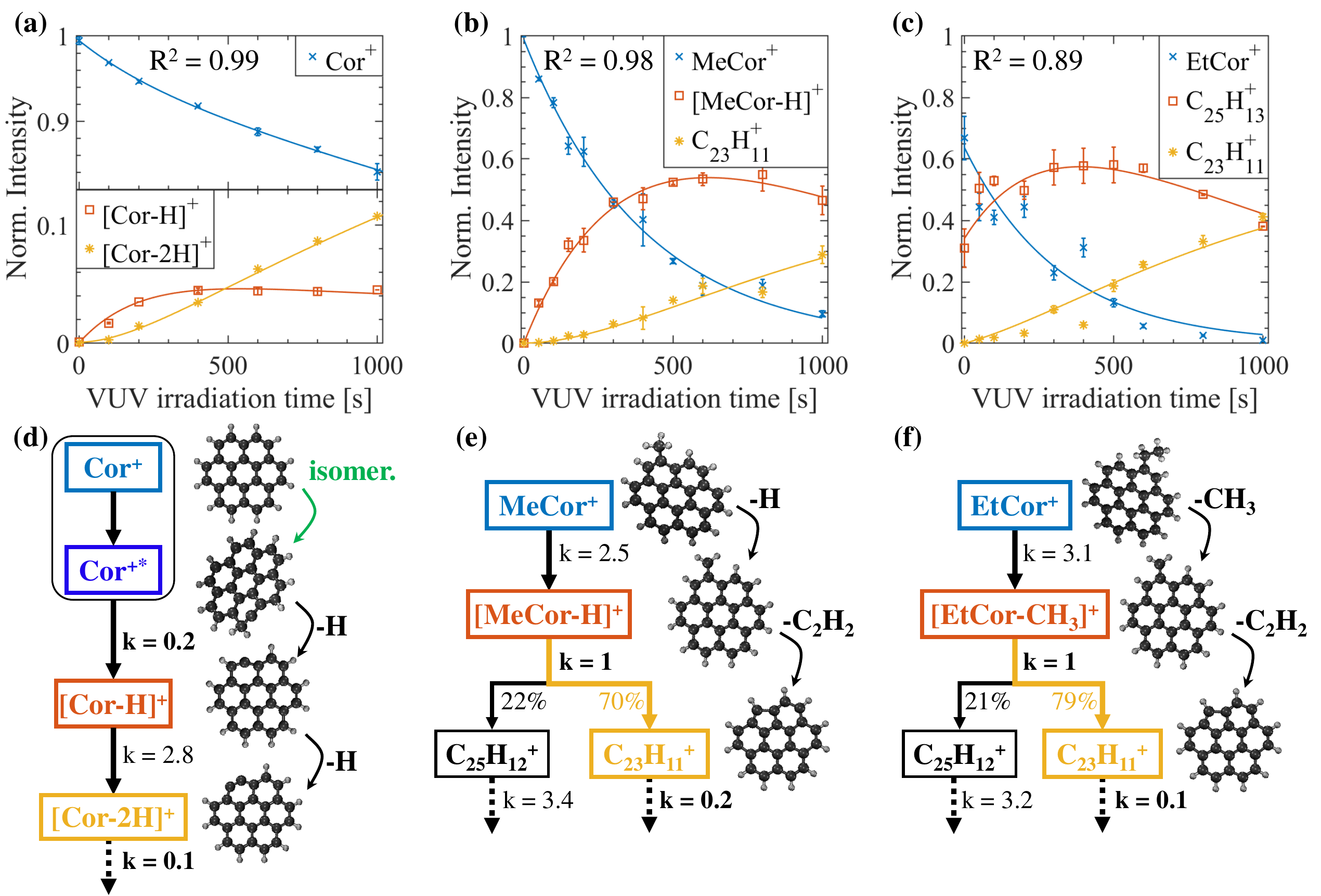}
    \caption{(a-c) VUV photoprocessing kinetic data (experimental points and fitting curves from the kinetic model) of the parents and main fragmentation channels of Cor$^+$, MeCor$^+$ and EtCor$^+$ and (d-f) their associated simplified fragmentation maps. We display the fragmentation rates ($\mathrm{k}$, in $\times 10^{-3}$s$^{-1}$) and branching ratios extracted from the fit performed with the kinetic model.}
    \label{fig:PIRENEA_results_kinetics_maps}
\end{figure*}

We studied the fragmentation kinetics of the coronene cation (\ce{C24H12+}, Cor$^+$) and its alkylated derivatives, namely methylcoronene (\ce{C25H14+}, MeCor$^+$), and ethylcoronene (\ce{C26H16+}, EtCor$^+$). 
The kinetic curves of the parents and main fragmentation channels of these three PAHs are shown in Figure~\ref{fig:PIRENEA_results_kinetics_maps}(a-c). We observe that the fragmentation of the bare Cor$^+$ is much less efficient than its methylated or ethylated derivatives. The kinetic curves are analyzed with a simple population/depopulation model whose solution fits the data (full line and R$^2$ in Figure~\ref{fig:PIRENEA_results_kinetics_maps}(a-c)). From this model, we derive a fragmentation map with fragmentation rates ($\mathrm{k_{frag}}$) and branching ratios for the different fragmentation channels. We show a simplified version of these maps in Figure~\ref{fig:PIRENEA_results_kinetics_maps}(d-f) for Cor$^+$, MeCor$^+$ and EtCor$^+$, respectively. They illustrate well the difference of fragmentation pathways between a bare PAH and its alkylated derivatives. Interestingly, we observe that the first H-loss step in MeCor$^+$ is mimicked by a CH$_3$-loss in EtCor$^+$. It shows that similar photoproducts are involved in the fragmentation cascades of these alkylated PAHs.

\begin{figure}[!ht] 
    \centering
    \includegraphics[width=0.6\textwidth]{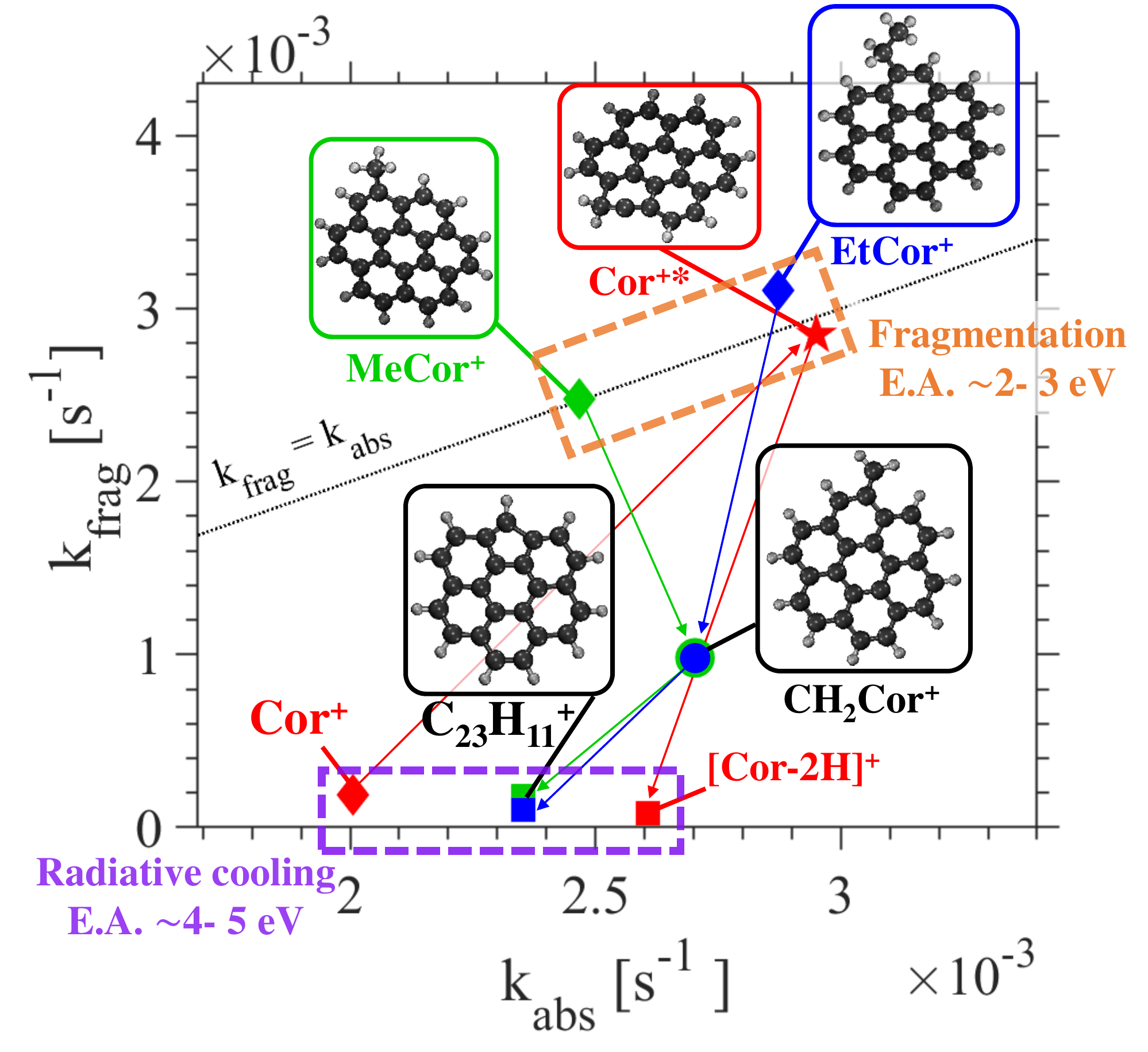}
    \caption{Fragmentation rates as a function of theoretical (TD-DFT) photoabsorption rates for Cor$^+$, MeCor$^+$, EtCor$^+$, and their photofragments. The dotted line indicates species that dissociate for each absorbed VUV photon, while species far below have high radiative cooling rates. Adapted from Ref.\cite{marciniak2021}.    }
    \label{fig:PIRENEA_results_summary}
\end{figure}

Figure~\ref{fig:PIRENEA_results_summary} summarizes the VUV photostability of the studied PAHs and their fragments. Our results provide experimental support for the isomerization of Cor$^+$ upon VUV irradiation, which is presumably due to the migration of an H atom \cite{trinquier2017_H_shifted_isomers} (e.g see Figure~\ref{fig:PIRENEA_results_kinetics_maps}(d), see also discussion by A.G.G.M. Tielens in this volume). In addition, Cor$^+$ is found to be very stable compared to MeCor$^+$ or EtCor$^+$ which have higher fragmentation rates and exhibit carbon loss channels. However, the VUV fragmentation cascade of alkylated PAHs leads to the formation of a fragment bearing a peripheral pentagonal ring (\ce{C23H11+}), which is as stable as Cor$^+$ or [Cor-2H]$^+$. All of the latter species have a higher radiative cooling fraction (see Figure~\ref{fig:PIRENEA_results_summary}) and thus a higher probability of survival in PDRs.

This work supports a scenario in which the carriers of the 3.4\,$\mathrm{\mu}$m AIB are efficiently photo-dissociated and this leads to the production of small hydrocarbons in PDRs \cite{pety2005, pilleri2015}. This process results in the formation of PAHs containing a pentagonal ring, which can be a step in the formation of fullerenes \cite{berne2012}. Finally, our study provides evidence for isomerization processes which could result in the formation of aliphatic C-H bonds in PAHs and provide an additional contribution to the 3.4\,$\mathrm{\mu}$m AIB \cite{jolibois2005}.

\section{Towards a study of gas-nanograin-photon interactions}

\begin{figure}[!ht] 
    \centering
    \includegraphics[width=1\textwidth]{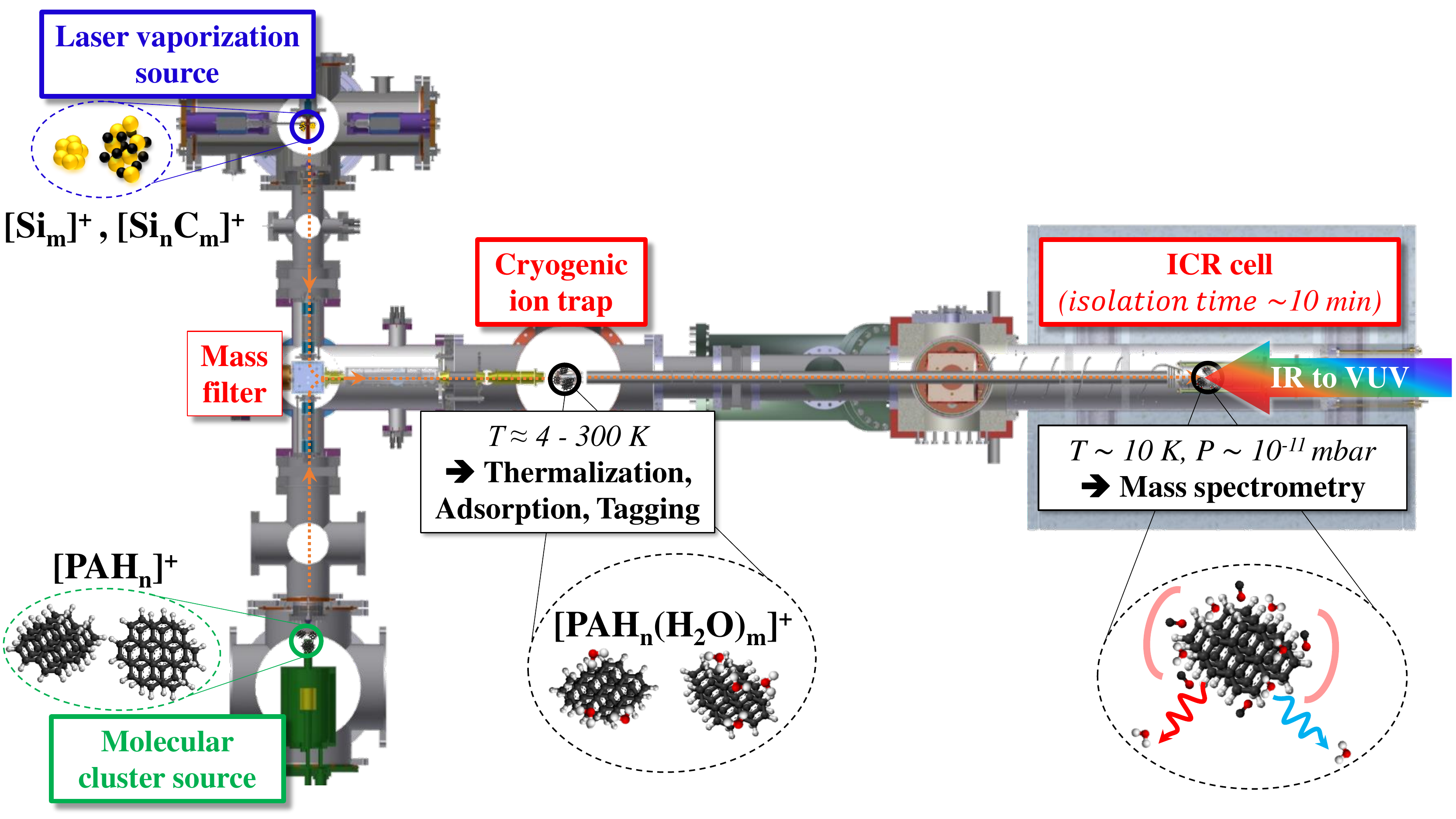}
    \caption{Scheme of PIRENEA 2 with its most important features and illustrations of the possible gas-nanograin-photon interactions which are planned to be studied.}
    \label{fig:PIRENEA2_setup}
\end{figure}

Upcoming observations from the James Webb Space Telescope (JWST) will allow us to study, with unprecedented sensitivity and precision, the chemical evolution of matter in PDRs, including that of AIB carriers. We expect to better characterize PAH precursors in PDRs, the so-called evaporating very small grains \cite{pilleri2012}.  We also expect to detect nano-sized species, including molecular clusters, and their interaction with gas phase molecules through adsorption and desorption processes. In this context, PIRENEA 2 has been designed to generate and study a large panel of species (see Figure~\ref{fig:PIRENEA2_setup}, \cite{bonnamy2018}). The setup combines two versatile nanograin sources, namely (i) a laser vaporization source to produce species of interest for stardust chemistry \cite{Berard2021} and (ii)  a source of molecular clusters to produce homogenous or heterogenous weakly bounded clusters (e.g. PAH$_n^+$ or PAH$_n^+$(H$_2$O)$_m^+$) \cite{zamith_thermal_2019,zamith2022}.
PIRENEA~2 includes a number of features that are or will be soon commissioned: (i) a post-source mass selection, (ii) a cryogenic electrostatic trap to perform adsorption of small molecules (e.g. CO or H$_2$O) or tagging with rare gas atoms (e.g. He or Ne) and (iii) a transfer and storage in the cryogenic ICR cell (similar to PIRENEA but with increased sensitivity). This will allow us to study photoinduced mechanisms at different photon energies. We plan to perform vibrational spectroscopy, including the characterization of C-C and C-H modes in relation to AIBs, and explore, by UV-VUV photoprocessing, the fragmentation and reactivity of a variety of nanograins containing carbon. 

\section{Conclusion and outlook}
PIRENEA provides the necessary environment to study the photophysics of PAHs under VUV irradiation in physical conditions that are relevant to PDRs. The extracted fragmentation rates and maps are useful for astrochemical models. In the near future, PIRENEA 2 will allow us to study the properties of isolated nanograins of interest for the interpretation of JWST observations and for astrochemistry in general. From a more fundamental point of view, we wish to study the role of intermolecular interactions in the energy redistribution of photoinduced processes.

\paragraph{Aknowledgments.}
This research has received funding from the European Research Council (ERC-2013-SyG, Grant agreement N$^{o}$610256 NANOCOSMOS).

\bibliography{Proceedings_ECLA2020_A_Marciniak}

\end{document}